\documentclass[aps,pre,reprint,groupedaddress]{revtex4-1}
\usepackage{amssymb,amsfonts,amsmath}
\usepackage[dvips]{graphicx,color,psfrag}

\begin{document}


\title{
Power-law exponent in multiplicative
Langevin equation with temporally correlated noise
}
\author{Satoru Morita}
\email[]{morita.satoru@shizuoka.ac.jp}
\affiliation{Department of Mathematical and Systems Engineering, Shizuoka University, Hamamatsu, 432-8561, Japan}
\date{\today}

\begin{abstract}
Power-law distributions are ubiquitous in nature.
Random multiplicative processes are a basic model 
for the generation of  power-law distributions.
It is known that, for discrete-time systems,
the power-law exponent decreases as the autocorrelation time of
the multiplier increases. 
However, for continuous-time systems,  
it has not yet been elucidated as to 
how the temporal correlation affects 
the power-law behavior.
Herein, we have analytically investigated a multiplicative Langevin equation with colored noise.
We show that the power-law exponent depends on the details of the 
multiplicative noise, in contrast to the case of  discrete-time systems.
\end{abstract}

\pacs{05.40.-a, 02.50.-r, 87.10.-e, 89.65.Gh}

\maketitle

\section{Introduction}

Fluctuations following power-law distributions have been
found not only in natural systems but also in social systems \cite{newman05,clauset,gabaix09}.
For instance, 
city sizes \cite{gabaix99,ioannides}, 
firm sizes \cite{ramsden,axtell}, 
stock returns \cite{mandelbrot,gabaix03} and
personal incomes \cite{champernowne,reed}
follow the power-law distribution 
\begin{equation}
P(x)\propto x^{-\gamma-1}.
\label{eq1}
\end{equation}
over large scales. 
This expression is widely known as Pareto's law \cite{pareto} or 
Zipf's law \cite{zipf}, 
and it has been investigated using various models.
A well-known mechanism that generates power-law distributions is the random multiplicative process
\cite{render,kesten,takayasu,levy,sornette,sornette98,biham,nakao,sato}. 
For example, consider the case of personal income:
if a person invests his/her money, he/she will get 
a certain percentage return that varies over time.
With repeated investments, the evolution of income approaches
\begin{equation}
x(t+1)=\lambda(t)x(t)+b,
\label{eqN0}
\end{equation}
where a small positive term $b$ is added 
to introduce a lower bound on the value of $x(t)$.
Here, the multiplier $\lambda(t)$ is a stochastic variable. 
If $\langle \log \lambda \rangle<0$ and $\lambda(t)$
sometimes takes values larger than one, 
the asymptotic distribution of $x(t)$ has a power law tail
\cite{takayasu,levy,sornette}.
Although the added term $b$ had been a stochastic variable in the previous works\cite{takayasu,levy,sornette},
we have set $b$ as a constant for this investigation 
because fluctuations of $b$ 
will not affect the power-law tail of the distribution 
when it is sufficiently small.
If the multiplier is uncorrelated, the power-law exponent 
is given by a solution of  eq.~(\ref{eqN2}) \cite{kesten,takayasu}.
\begin{equation}
\langle \lambda^{\gamma} \rangle=1
\label{eqN2}.
\end{equation}
In the case that the variance of $\log \lambda$ exists,
the power-law exponent is calculated
approximately by eq.~(\ref{eqN1}) \cite{sornette,sornette98}.
\begin{equation}
\gamma=\frac{|\langle\log\lambda\rangle|}
{\langle(\log\lambda)^2\rangle-\langle\log\lambda\rangle^2}.
\label{eqN1}
\end{equation}
If the multiplier is temporally correlated,
the power-law exponent $\gamma$ decreases as 
the autocorrelation time increases \cite{sato,morita}.
This relation can be explained intuitively by pointing out that
the temporal correlation of the multiplier tends to influence
the size of its fluctuations, i.e., the denominator of 
eq.~(\ref{eqN1}).

In contrast, for continuous-time systems, 
the effect of a temporally correlated multiplier 
on the power-law exponent has not yet been
investigated sufficiently.
This may be because 
results for discrete-time systems can be applied directly
for continuous-time systems.
In this paper, we investigate the relationship 
between the power-law exponent and 
the temporal correlation of the multiplier.
First, we introduce a
continuous-time version of eq.~(\ref{eqN0}).
Next, we analytically estimate the power-law 
exponent for this model
on the condition that the autocorrelation time is small. 
Finally, we perform numerical simulations to confirm
the predictions.

\section{Model}

We consider a Langevin equation (stochastic differential equation)
\begin{equation}
\frac{dx(t)}{dt}=(-r+\xi(t))x(t)+\epsilon,
\label{eq2}
\end{equation}
where $r$ and $\epsilon$ are positive constants, and 
$\xi(t)$ is a temporally correlated noise term
whose mean value is zero. 
Since $-r+\xi(t)$ represents the growth rate, $-r$ is the mean growth rate and
$\xi(t)$ is the deviation from the mean.
Here, we have assumed that $\xi(t)$ is characterized by 
intensity $D$ and autocorrelation time $\tau$,
where the autocorrelation function of $\xi(t)$ is given as  
\begin{equation}
\langle \xi(t)\xi(t')\rangle=
\frac{D}{\tau}e^{\textendash|t-t'|/\tau}.
\label{eq3}
\end{equation}
Thus, the power spectrum of  $\xi(t)$ is
\begin{equation}
S(\omega)=\frac{2D}{\tau^2 \omega^2+1}.
\end{equation}
For $\omega=0$, $S(0)=\int_{-\infty}^{\infty} \xi(t)^2
dt=2D$.
In the limit of $\tau\to 0$,   
eq.~(\ref{eq3}) converges to $2D \delta(t-t')$,
which indicates that $\xi(t)$ becomes white noise.
For these reasons, $D$ is referred to as the noise intensity
\cite{sancho,hanggi,patil}.

If we neglect quadratic and higher order terms
for infinitesimally small $\Delta t$, the
Langevin equation in (\ref{eq2}) can be rewritten as 
\begin{equation}
x(t+\Delta t)=[1+(-r+\xi(t))\Delta t]x(t)+\epsilon \Delta t .
\label{eq4}
\end{equation}
If we substitute $\lambda(t)=1+(-r+\xi(t))\Delta t$ and $b=\epsilon \Delta t$,
eq.~(\ref{eqN0}) is equivalent to eq.~(\ref{eq4}).
Note that for eq.~(\ref{eq4}), the noise term is proportional to
$\Delta t$, in contrast that
noise term is proportional to $\sqrt{\Delta t}$ for
usual stochastic differential equations.
Consequently, in calculating Eq.~(\ref{eqN0}),
there is no problem which calculus, Ito or Stratonovich, is used \cite{patil}.

The noise term that satisfies condition (\ref{eq3})
is not determined uniquely. 
Herein, we focus on a simple case.
We assume that  $\xi(t)$ has a stationary distribution $\rho(\xi)$. 
The value of $\xi(t)$ changes at each point in time and
remains constant over the subsequent moments.
The points in time follow the Poisson process with 
a rate of $1/\tau$.
At each point in time, a new value of $\xi(t)$ is chosen at random 
from $\rho(\xi)$.
Consider the case in which $\rho(\xi)$ is an
$n$-point discrete distribution, i.e.,
$\xi$ is $\xi_i$ with probability $\rho_i$ for $i=1,2,\dots,n$.
In this case, the evolution of the probability $p(\xi_i,t)$
of  $\xi=\xi_i$ at time $t$
is described by 
\begin{equation}
\frac{d}{dt}
\left(
\begin{array}{cc}
p(\xi_1,t)\\
p(\xi_2,t)\\
\\
p(\xi_n,t)
\end{array}
\right)=\frac{1}{\tau}
A 
\left( \begin{array}{cc}
p(\xi_1,t)\\
p(\xi_2,t)\\
\\
p(\xi_n,t)
\end{array}
\right).
\label{eqN3}
\end{equation}
Here, $\frac{1}{\tau}A$ represents
the transition rate matrix; the matrix $A$ is defined as
\begin{equation}
A_{ij}=-\delta_{ij}+\rho_i ,
\end{equation}
where $\delta_{ij}$ is Kronecker's delta.
Eq.~(\ref{eqN3}) 
delivers the correlation function (\ref{eq3})
from the fact that 
the dominant eigenvalue of $A$ is $0$ 
and all the rest eigenvalues are $-1$.

By neglecting the quadratic and higher order terms of $\Delta t$,
we obtain a discrete-time version of eq.~(\ref{eqN3})
\begin{equation}
\left(
\begin{array}{cc}
p(\xi_1,t+\Delta t)\\
p(\xi_2,t+\Delta t)\\
\\
p(\xi_n,t+\Delta t)
\end{array}
\right)=B
\left(\begin{array}{cc}
p(\xi_1,t)\\
p(\xi_2,t)\\
\\
p(\xi_n,t)
\end{array}
\right),
\end{equation}
where  $B$ represents the transition matrix, which is given as 
\begin{equation}
\begin{array}{lcl}
B_{ij}&=&\displaystyle \delta_{ij}+\frac{\Delta t}{\tau}A_{ij}\\
&=&\displaystyle \left(1-\frac{\Delta t}{\tau}\right)\delta_{ij}+\frac{\Delta t}{\tau}\rho_i.
\end{array}
\end{equation}
If $\rho(\xi)$ is a continuous distribution, then
the operators $A$ and $B$ can be defined similarly.  

\section{Calculation of power-law exponent}

The stationary distribution of $x$ for 
stochastic process described by eq.~(\ref{eq2}) or (\ref{eq4}) 
has a power-law tail.
Let $\bar{x}^{\alpha}(\xi,t)$ be defined as  
the expected value of $x^{\alpha}(t)$ on the condition that $\xi(t)=\xi$.
Thus, the expected value of $x^{\alpha}(t)$ is given as
\begin{equation}
\bar{x}^{\alpha}(t)=\int \bar{x}^{\alpha}(\xi,t)\rho(\xi).
\end{equation}
The power-law exponent $\gamma$ is determined by
the boundary between growth and decay of $\bar{x}(t)^{\alpha}$ \cite{kesten,takayasu}.
In the case of $n$-point discrete distribution, neglecting the added term 
$\epsilon$, we have
\begin{equation}
\left(
\begin{array}{cc}
\bar{x}^{\alpha}(\xi_1,t+\Delta t)\\
\bar{x}^{\alpha}(\xi_2,t+\Delta t)\\
\\
\bar{x}^{\alpha}(\xi_n,t+\Delta t)
\end{array}
\right)=BC
\left(\begin{array}{cc}
\bar{x}^{\alpha}(\xi_1,t)\\
\bar{x}^{\alpha}(\xi_2,t)\\
\\
\bar{x}^{\alpha}(\xi_n,t)
\end{array}
\right),
\end{equation}
where the matrix $C$ is defined as
\begin{equation}
C_{ij}=\left[1+\Delta t(-r+\xi_i)\right]^{\alpha}\delta_{ij}.
\end{equation}
As a result, the growth rate of $\bar{x}(t)^{\alpha}$ is given by the
dominant eigenvalue of the matrix $BC$.

Thus, the power-law exponent $\gamma$
can be derived by the condition that the
eigenvalue of $BC$ equals one, i.e.,
\begin{equation}
\det(BC-I)=0, 
\label{eqN5}
\end{equation}
where $I$ is the unit matrix.
Neglecting the quadratic and higher order terms of $\Delta t$ again,
we obtain
\begin{equation}
[BC-I]_{ij}=\Delta t \left\{\left[(-r+\xi_i)\alpha-\frac{1}{\tau}
\right]\delta_{ij}+\frac{1}{\tau}\rho_{i}\right\}.
\end{equation}
By simple algebra, the equation can be rearranged as
\begin{equation}
\begin{array}{cc}
\det(BC-I)=&
\displaystyle \Delta t^n 
\left[1+\sum_{i=1}^n\frac{\rho_i}{\tau(-r+\xi_i)\alpha-1}\right]
\\
&\times \displaystyle \prod_{i=1}^n
\left[(-r+\xi_i)\alpha-1/\tau\right].
\end{array}
\end{equation}
Consequently, 
the power-law exponent $\gamma$ can be determined by 
solving eq.~(\ref{eq14}).  
\begin{equation}
\sum_{i=1}^n\frac{\rho_i}{\tau(-r+\xi_i)\gamma-1}=-1.
\label{eq14}
\end{equation}
Expanding the result of eq.~(\ref{eq14}) to the case of 
a continuous distribution $\rho(\xi)$, we obtain
\begin{equation}
\int\frac{\rho(\xi)}{\tau(-r+\xi)\gamma-1}d\xi=-1.
\label{eq15}
\end{equation}

From eq.~(\ref{eq3}), the variance of $\rho(\xi)$ is $D/\tau$.
Using the rescaled variable $\xi'=\xi/\sqrt{D/\tau}$,
whose distribution is 
\begin{equation}
\rho'(\xi')=\rho(\xi'\sqrt{D/\tau})\sqrt{D/\tau},
\end{equation}
eq.~(\ref{eq14}) can be rewritten as
\begin{equation}
\int\frac{\rho'(\xi')}{(-r\tau+\xi'\sqrt{D\tau})\gamma-1}d\xi'=-1.
\label{eq17}
\end{equation}
Here, the variance of $\rho'(\xi)$ is one.
Performing Maclaurin series expansion with respect to 
$\tau^{1/2}$ to the left-hand side of  eq.~(\ref{eq17}),
the condition of eq.~(\ref{eq17}) can be rewritten as
\begin{equation}
\begin{array}{l}
(r\gamma-D \gamma^2)-S D^{3/2}\gamma^3 \tau^{1/2}
\\
-\gamma^2\left(r^2-3r\gamma D+\gamma^2 D^2(K+3)\right)\tau+O(\tau^{3/2})=0,
\end{array}
\label{eq18}
\end{equation}
where $S$ and $K$ are the skewness and 
the kurtosis of $\rho(\xi)$, respectively.
\begin{eqnarray}
S&=&\displaystyle \langle\xi^3\rangle/\langle\xi^2\rangle^{3/2}, \\
K&=&\displaystyle \langle\xi^4\rangle/\langle\xi^2\rangle^{2}-3.
\end{eqnarray}
In the limit $\tau\to 0$ (the white noise limit), 
the first term on the left-hand side of 
(\ref{eq18}) must be zero, so that we can obtain
\begin{equation}
\gamma=\frac{r}{D}.
\label{eqNN}
\end{equation}
Eq.~(\ref{eqNN}) coincides with the result of the previous 
work \cite{nakao} in which the multiplier was the white noise
for Stratonovich stochastic differential equation.

Consider the case of $\tau\ll1$.
In the case when $\rho(\xi)$ is asymmetric ($S\neq 0$),
we obtain an approximation formula (\ref{eq24}), 
taking the first and second terms of of the left-hand side of 
(\ref{eq18}) into account.
\begin{equation}
\gamma\simeq \frac{r}{D}\left(1-rS\sqrt{\frac{\tau}{D}}\right).
\label{eq24}
\end{equation}
Since the leading term of (\ref{eq24}) is the square root of $\tau$,
the skewness $S$ seriously affects
the dependence of the power-law exponent $\gamma$
on the autocorrelation time $\tau$.
If $S<0$, the power-law exponent $\gamma$ increases quickly with 
the autocorrelation time $\tau$, whereas if  $S>0$, the opposite is true.

In the case when $\rho(\xi)$ is symmetric ($S=0$),
we obtain another approximation formula,
while taking the linear term of $\tau$ and the constant term into account
\begin{equation}
\gamma\simeq \frac{r}{D}\left[1-r^2(K+1)\frac{\tau}{D}\right].
\end{equation}
If $K>-1$, the power-law exponent $\gamma$ decreases when 
the autocorrelation time $\tau$ increases. If $K<-1$, the opposite is true. 

\begin{figure}[tb]
\includegraphics[width=8.6cm]{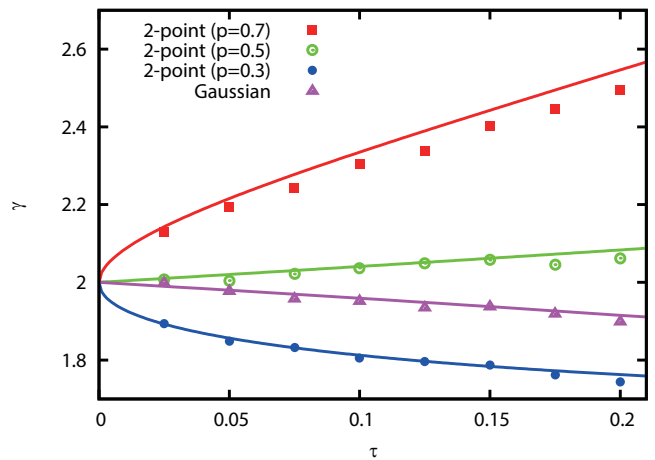}
\caption{(Color online) 
Power exponent $\gamma$ is plotted as
a function of the autocorrelation time $\tau$,
when the stationary distribution $\rho(\xi)$ 
of the multiplier two-point discrete distributions (\ref{eq2PD}) 
for $p=0.7, 0.5, 0.3$
($S=-0.87\cdots, 0, 0.87\cdots$)
and  a Gaussian distribution (\ref{eqGD}).
The other parameters are set to $r=0.1$, $\epsilon=0.0001$ and $D=0.05$.
The curves represent the
theoretical calculations and symbol makers 
represent numerical simulations.
To estimate $\gamma$ numerically,
we considered ensembles with $10^6$ elements that follow (\ref{eq4})
for $\Delta t=0.001$, and 
employed the maximum-likelihood method\cite{newman05} 
for the top $9\times10^5$ elements.
Each symbol represents an average over 50 samples,
where the size of the symbol is larger than the standard error.}
\label{fig1}
\end{figure}

\section{Examples}

Next, we examine 
two simple examples of the distribution $\rho(\xi)$.
First,  consider the case of two-point discrete distribution:
\begin{equation}
\xi=\left\{
\begin{array}{ccc}
\displaystyle \sqrt{\frac{D}{\tau}}\sqrt{\frac{1-p}{p}}&&(\mbox{probability} \ p)\\
\displaystyle -\sqrt{\frac{D}{\tau}}\sqrt{\frac{p}{1-p}}&&(\mbox{probability} \ 1-p)
\label{eq2PD}
\end{array}
\right. .
\end{equation}
In this case,
\begin{equation}
S=\frac{1-2p}{\sqrt{p(1-p)}} \ \mbox{and} \ K=\frac{1}{p(1-p)}-6. 
\end{equation}
By solving eq.~(\ref{eq14}) analytically, 
the power-law exponent can be expressed as follows:
\begin{equation}
\gamma=\frac{r}{D+Sr\sqrt{D\tau}-r^2\tau}.
\label{eq23}
\end{equation}
In fig.~1, curves for eq.~(\ref{eq23}) are plotted for 
$p=0.7, 0.5$, and $0.3$ ($S=-0.87\cdots, 0, 0.87\cdots$, respectively)
when we set $r=0.1$ and $D=0.05$.
If $S\neq0$, the curves have a parabola-like shape.
On the contrary, if $S=0$ ($p=0.5$),  
the power-law exponent $\gamma$ increases almost linearly with 
the autocorrelation time $\tau$, because $K=-2$.
These results agree with the predictions we made above.
To confirm our analytical results,
we perform numerical simulations of the Langevin equation (\ref{eq2}). 
Here, we apply the Euler method (eq.~(\ref{eq4})) for $\Delta t=0.001$.
Fig. 1 shows the consistency of numerical results 
with the analytical results.


Second, consider the case of Gaussian distribution
for the stationary distribution
\begin{equation}
\rho(\xi)=\frac{1}{\sqrt{2\pi D/\tau}}\exp
\left(-\frac{\xi^2}{2D/\tau}\right)
\label{eqGD}
\end{equation}
In this case, we cannot obtain an explicit expression for $\gamma$.
In fig.~1, the curve is plotted by calculating 
eq.~(\ref{eq17}) numerically.
The power-law exponent $\gamma$ decreases almost linearly 
as the autocorrelation time $\tau$ increases, because $S=0$ and $K=0$
for the Gaussian distribution.
Also in this case, the numerical results are consistent 
with the analytical results (see the triangles markers in fig.~1).

\section{Conclusion}

In summary, we showed that
the power-law exponent $\gamma$ 
for a stochastic differential equation
depends on the stationary distribution of 
the multiplier term 
even if the autocorrelation function is the same.
Particularly, in the case when 
the skewness $S$ of the stationary distribution 
is nonzero, a slight change to the autocorrelation time can
have a dramatic effect on the power-law exponent $\gamma$.
If $S=0$, the relation between 
the power-law exponent and the autocorrelation time 
is determined by whether  kurtosis $K$ is larger than $-1$ or not.  
In practice, for continuous distributions
which have the same tails as the Gaussian distribution
or longer tails ($K\geq 0$), 
the power-law exponent $\gamma$ would decrease gently
as the autocorrelation time increases.
For example, empirical works 
for the sales of American companies \cite{stanley} 
and the national GDPs \cite{lee} reported that 
the growth rates follow 
symmetric exponential distributions ($S=0$ and $K=3$).
Another report finds that 
the growth rates for the income of Japanese companies
follow an asymmetric exponential distribution ($S<0$) \cite{mizuno}.
The latter case is very interesting because the temporal correlation
may significantly affect power-law behavior.
Future works will need to address practical data
to further explore this topic. 

Our results are seemingly inconsistent with
previous studies reporting that
the power-exponent is proportional to the inverse of the
autocorrelation time $\tau$ for large values of $\tau$ \cite{sato,morita}.
These studies
assumed that the autocorrelation function can be described with
\begin{equation}
\langle \xi(t)\xi(t')\rangle=D e^{\textendash|t-t'|/\tau}
\label{eqN6}
\end{equation}
instead of eq.~(\ref{eq3}).
Although eq.~(\ref{eqN6}) is suitable for discrete-time systems,
eq.~(\ref{eq3}) is more appropriate for 
describing continuous-time systems
because $\int_{-\infty}^{\infty} \xi(t)^2 dt=2D$ 
for eq.~(\ref{eq3}) 
and the autocorrelation function converges to $2D \delta(t-t')$
in the limit of $\tau\to 0$, as is mentioned above.
On the other hand, eq.~(\ref{eqN6}) 
cannot converges to white noise in the limit $\tau\to 0$.

Finally it should be noted that we have focused on 
a simple case that satisfies eq.~(\ref{eq3}).
Generally, for an operator $A$ 
when the dominant eigenvalue is $0$ 
and all the other eigenvalues are $-1$,
eq.~(\ref{eqN3}) produces noise 
with an exponential autocorrelation function (\ref{eq3}).
In this case, the power-law exponent $\gamma$
can be calculated by solving eq.~(\ref{eqN5}) at least in principle.
An alternative method used to
generate temporally correlated noise
is the Ornstein-Uhlenbeck process
\begin{equation}
d\xi(t)=-\frac{1}{\tau}\xi(t)dt+\frac{\sqrt{2D}}{\tau}dW(t),
\end{equation}
where $W(t)$ denotes the Wiener process.
In this case, the operator $A$ is given by using the Fokker-Planck equation.
However, the derivation of the operator $B$ and
solution for  eq.~(\ref{eqN5}) are quite difficult.
Calculating the power-law exponent 
for such cases remains an open problem to be addressed 
in future work.

\begin{acknowledgments}
This work was supported by CREST, JST.
\end{acknowledgments}

\end{document}